\documentclass[preprint,proceedings]{rmaa}

\suppressfulladdresses 

\usepackage{paralist}

\usepackage{psfrag,color}

\usepackage{natbib}

\newcommand{\chandra}{{\it CHANDRA}}

\newcommand{\suzaku}{{\it Suzaku}}
\newcommand{\xmm}{{\it XMM}}

\newcommand{\ec}{$\eta$ Carinae}

\SetYear{2006}
\SetConfTitle{ Circumstellar Media and Late Stages of Massive Stellar Evolution}

\title{Eta Car and Its Surroundings: the X-ray Diagnosis} 

\author{
 M. F. Corcoran\altaffilmark{1,2} \& Kenji Hamaguchi\altaffilmark{1,2}}

\altaffiltext{1}{CRESST and X-ray Astrophysics Laboratory NASA/GSFC, Greenbelt, MD 20771, USA. (corcoran@milkyway.gsfc.nasa.gov).}

\altaffiltext{2}{Universities Space Research Association, 10211 Wincopin Circle, Suite 500 Columbia, MD 21044, USA.}

\shortauthor{Corcoran \& Hamaguchi}
\shorttitle{Eta Car and Its Surroundings}

\listofauthors{M. F. Corcoran, K. Hamaguchi}
\indexauthor{Corcoran, M. F., Hamaguchi, K.}

\abstract{X-ray emission from the supermassive star Eta Carinae (\ec) originates from hot shocked gas produced by current stellar mass loss as well as ejecta from prior eruptive events.  Absorption of this emission by cool material allows the determination of the spatial and temporal distribution of this material.  Emission from the shocked gas can provide important information about abundances through the study of thermal X-ray line emission.  We discuss how studies of the X-ray emission from Eta Car at a variety of temporal, spatial and spectral scales and resolutions have helped refine our knowledge of both the continuous and discrete mass loss from the system, and its interactions with more extended material around the star.
}

\addkeyword{Stars: Eta Car}
\addkeyword{ISM: Jets and outflows}
\addkeyword{Stars: Binaries}
\addkeyword{Stars: Mass loss}

\begin{document}
\maketitle

\section{Unwrapping the Riddle}
\label{sec:intro}

Churchill's well-quoted epigram on the Soviet Union is also appropriate for the massive, extremely luminous star \ec\ \citep{1997ARA&A..35....1D}: the star is a riddle, wrapped in a mystery, surrounded by enigma.  In the riddle of \ec, the mystery is the Homunculus nebula \citep{1950ApJ...111..408G, 1843MNRAS...6....9M} which was ejected in an uncertain manner during and after the ``Great Eruption'' of the star starting in the mid-1800's \citep{1838MNRAS...4..121H}.  The enigma is the Great Carina Nebula, a dusty system of massive molecular clouds which provides the raw material for star formation in this region and which is ionized and shaped by radiation and stellar winds from those selfsame stars.

Thus the Carina Nebula exemplifies the fundamental ``feedback loop''  involved in the process of star formation: massive clouds which organize themselves by gravity into massive stars, whose subsequent radiative fields and mass ejections (and supernovae) either trigger the formation of additional new stars, or shut down the process altogether.  The presence of \ec, an unstable, luminous, periodically (and aperiodically?) variable, mass-losing binary star in this environment only serves to further highlight the complexities of this relation.

X-radiation is a good illuminator of these complex processes.  X-rays produced by massive stars (which, as in another well-quoted epigram, live fast, die young and leave good-looking -- or at least interesting -- corpses) provide probes of chemical evolution, duplicity, magnetic fields, shock heating, wind speeds, turbulent velocities, densities and other vital quantities. A source of hard X-ray emission in the Carina Nebula signals its presence to astronomers even through enormous ($\sim 10^{24}$ cm$^{-2}$) column depths of material in which it might be embedded, even if its X-ray luminosity is as low as $\approx10^{30}$ ergs s$^{-1}$ (which corresponds to an early B-type main sequence star or an active lower-mass star).  But large absorbing columns in front of embedded sources mean that important, low energy X-ray diagnostics (like the Ly-$\alpha$ lines of hydrogenic carbon, oxygen and nitrogen) are entirely absorbed. This is a particular problem in the case of \ec, since the star is buried behind the (optically) thick walls of the Homunculus. 

\section{From the Insight Out}
\label{sec:insight}

\begin{figure*}[!t]
  \includegraphics[width=2\columnwidth]{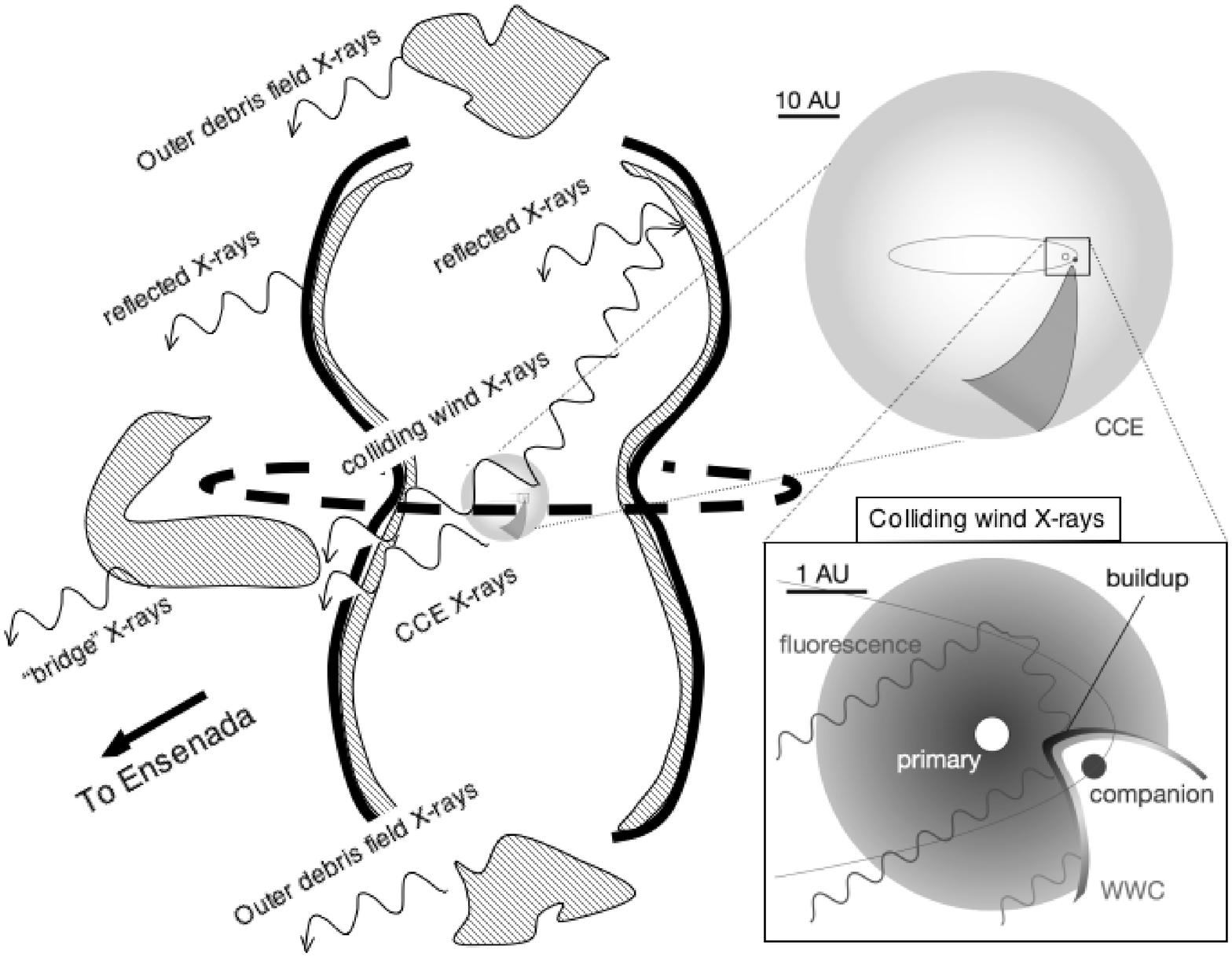}
  \caption{A cartoon representation of the \ec\ system showing the various X-ray emission and absorption components. Adapted from \citet{kenji}. Variable, hot ($kT\approx4$ keV)  X-ray emission is generated at the wind-wind interface in the \ec\ binary system deep inside the Homunculus. These X-rays are absorbed and scattered by the wind of \ec~A.  X-rays from the wind-wind collision also reflect off the cold material in the walls of the Homunculus and appears as a hot, heavily absorbed emission.  Cooler emission ($kT\approx 1$ keV), probably extended, resides within $1''$ of the binary. Even cooler emission ($kT\approx0.6$ keV) from the collision of the outer ejecta (in the outer debris field) with the pre-existing circumstellar medium surround the entire system. Absorption is due to the wind of \ec~A, material in the Homunculus, material just beyond the Homunculus, and the ISM. The entire system is embedded in the soft, X-ray bright emission from the Carina Nebula.}
  \label{fig:cartoon}
\end{figure*}

Insight to the whole process, in (enormous) miniature, can be obtained by considering \ec\ itself.  Figure \ref{fig:cartoon} shows a cartoon of the \ec\ system (an ``$\eta$ Car-toon'') adapted from a much nicer figure in \citet{kenji}, based largely on an \xmm\ X-ray monitoring campaign on the star in mid-2003 when the star was experiencing one of its periodic ``X-ray Minima'' (which recur every 2024 days).  From right to left it works like so: Variable X-rays generated periodically by a wind-wind collision in an extremely eccentric, long (2024 day) period binary pass through the wind of \ec, revealing important characteristics of the wind through absorption and fluorescent emission. The ``colliding wind'' X-ray emission is produced by the collision of \ec's wind (with $\dot{M}\sim\mbox{ a few }\times 10^{-4} M_{\odot}$ yr$^{-1}$ -- or more -- and a terminal speed of $V_{\infty}\approx500$ km s$^{-1}$) with a fast thin wind of a companion star (with $\dot{M}\sim\mbox{ a few }\times 10^{-5} M_{\odot}$ yr$^{-1}$ and a terminal speed of $V_{\infty}\approx3000$ km s$^{-1}$). Estimates of the stellar parameters are given in Table \ref{tab:ecparams}.  These numbers are largely taken from \citet{hillier01}, \citet{corc01}, \citet{hotp},  \citet{katya05} and \citet{corc05}. 

A constant X-ray emission component (the ``CCE'', or central constant emission), only revealed during the brief X-ray minimum, arises somewhere near the binary, possibly (as suggested in the cartoon) due to the collision between the outflowing, shocked colliding wind gas with ambient gas and dust, which creates a ``hot bubble'' of shocked gas as the colliding wind interface sweeps around due to orbital motion. This constant component has an X-ray temperature of a keV or so, and shows thermal line emission indicative of shocked gas moving with a pre-shocked velocity of $\sim1000$ km s$^{-1}$. This is considerably faster than \ec's ``normal'' wind but in the ballpark of the velocity of the polar wind, and, probably, of the outflow speed of the shocked material in the wind-wind collision.  All these emission components finally are absorbed by the gas and dust in the Homunculus nebula far from the binary before being detected at earth.

\begin{table}[!t]\centering
  \setlength{\tabnotewidth}{\columnwidth}
  \tablecols{4}
  \caption{Estimated System Parameters for \ec} \label{tab:ecparams}
  \begin{tabular}{lccc}
    \toprule
    Parameter & \multicolumn{1}{l}{$\eta$ Car A} & \multicolumn{1}{l}{$\eta$ Car B} & \multicolumn{1}{l}{System}\\
    \midrule
  Mass M$_{\odot}$ & $90$ & $30$??\\
   Radius R$_{\odot}$  & $150$ & $20$??\\
   Lumin. $10^{6}$ L$_{\odot}$ & $4$ & $0.9$?\\
   $T_{eff}$ kK   & $15$ & $34$?\\
  $\dot{M}$ M$_{\odot}$/yr & $10^{-4}-10^{-3}$ & $10^{-5}$?\\
$V_{\infty}$ km/s   & $500-1000$ & $3000$?\\
Period (d) &  & & $2024\pm2$\\
$e$ &  & & $0.8-0.95$\\
$a$ AU &  & & 15?\\
$i^{\circ}$ &  & & $45-90$\\
    \bottomrule
   \end{tabular}
\end{table}

\section{Effects on the Circumstellar Environment}

The circumstellar material around \ec\ mostly consists of the detritus of the giant eruptions in 19$^{th}$ century. Very near the star the wind from the system must have cleared out a substantial volume.  The twin lobes of the Homunculus are believed to be  largely hollow, thin shells, \citep{2006ApJ...644.1151S} separated by a rather thick disk.  This disk is sometimes called the ``equatorial disk'' or ``equatorial skirt'', though it's not clear what equator is referred to, although it seems a pretty good bet that it's the rotational equator of \ec~A. In any event, even though the interiors of the lobes  are hollow, there's still quite a bit of junk near \ec.  Most notable are the ``Weigelt Blobs'' \citep{blob} rather thick condensations within $\sim200$ AU of the star.  

Orbital motion of the companion around \ec\ causes a variety of periodic (and perhaps quasi-periodic) variations in this material. Most significant are variations in the ionization state.  These changes are driven by the photospheric UV radiation of \ec~B (although X-ray emission produced in the wind-wind shock might be important in driving the local ionization state of the gas). The nature of the system is aspherical, and the wind from \ec~A prevents UV radiation from the companion from ionizing that portion of the circumstellar material on the far side of the star from \ec~B.  During periastron passage, the companion is basically engulfed in the wind of \ec\ and the ionized zone collapses, and formerly ionized material recombines.  This implicitly means that the wind from \ec\ is asymmetric as well: ionized on the side near the companion, less ionized on the opposite side.  The ionization state asymmetry may have significant effects on the driving of the wind, since the ``non-ionized'' wind has more ionic transitions to transfer momentum via line absorption from the radiation field of \ec\ to the gas. This is very similar to the ``after-burner'' effect suggested by Hillier in which recombination of low-ionization stages of iron in the outer wind of \ec\ could re-accelerate the wind (even in the absence of a hot companion star). 
Similarly, the portion of the wind which is shadowed from the companion's radiation might feel a stronger radiative acceleration away from the primary than the ionized side of the wind near the companion.  The UV flux from the companion might also help to decelerate the primary's wind near the companion.   Similarly, the portion of the wind which is shadowed from the companion's radiation might feel a stronger radiative acceleration than the ionized side of the wind near the companion.  Such a ``two-state wind'' would be bipolar in every sense of the word.  Interactions between the two winds would also distort the purely radial outflow of \ec's wind into a spiral, similar to the ``pinwheel nebula'' produced by massive interacting binaries, and traced by dust formation \citep[for example, ][]{2006Sci...313..935T}.  One wonders if such ``spiral density waves'' moving out from the central binary might have some connection to the discrete velocity components reported recently by \citet{2006ApJS..163..173G} from high spatial resolution HST/STIS spectra in the line of sight to the star, or whether such episodic dust formation near periastron passage accounts for significant IR variability \citep{2004MNRAS.352..447W}.

X-rays interact with the more distant material of course, and not simply by absorption.   X-rays from the central star (or, more properly, the central wind-wind collision) reflect off the cold material in the Homunculus.  This reflected emission has been resolved by the \chandra\ X-ray observatory \citep{2004ApJ...613..381C} during the brief X-ray minimum when the glare from the central source is reduced.  This emission offers a unique diagnostic tool since it provides, in principal, the first 3-d view of a colliding wind X-ray emitting region, similar to the reflected optical emission which gives a 3-d view of \ec's wind.  But this phenomenon is difficult to exploit with current technology, since it's of low surface brightness and it's difficult or nearly impossible to disentangle from the scattered emission from the central source, even with \chandra, for most of the orbit when the colliding wind source is bright.    

\begin{figure}[!t]
  \includegraphics[width=\columnwidth]{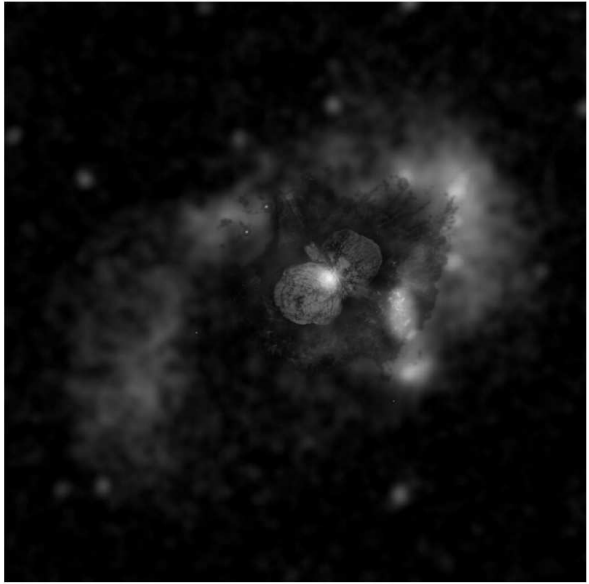}
  \caption{Comparison of a \chandra\ X-ray image of \ec\ and an HST/WFPC2 image of the Homunculus. The X-ray emission is extended and associated with the ``Outer Debris Field'' which surrounds the bipolar Homunculus nebula.  \ec\ is the optical- and X-ray-bright point at the center of the image.  WFPC2 image courtesy of N. Smith and J. Morse.}
  \label{fig:ecxo}
\end{figure}

\begin{figure*}[!t]
\centering
  \includegraphics[width=1.5\columnwidth]{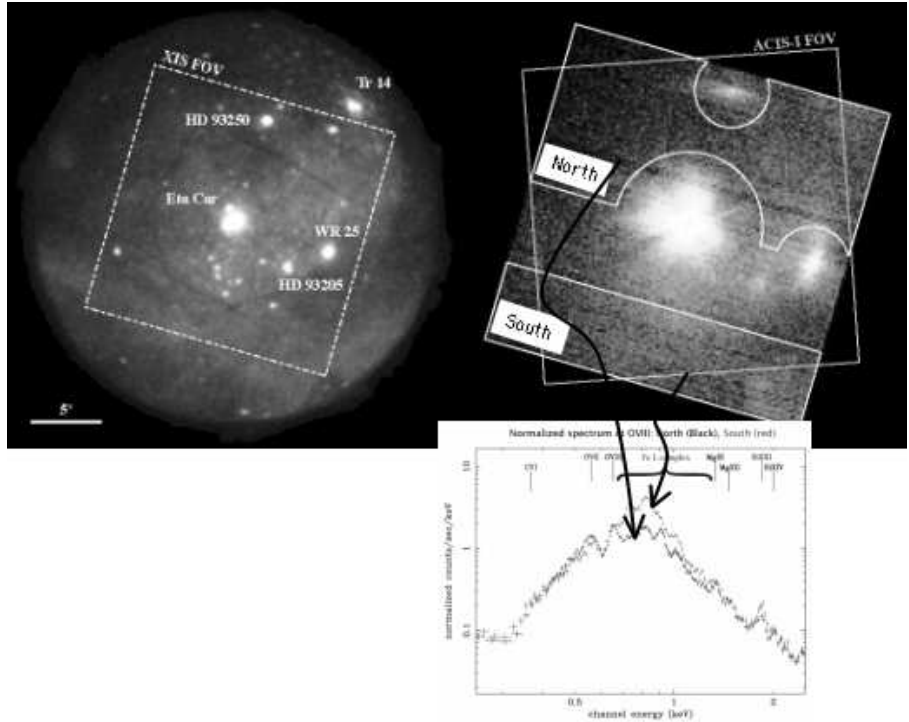}
  \caption{\textit{Left:} An \xmm\ X-ray image of the Carina Nebula. \textit{Right}: \suzaku\ X-ray image of the Carina Nebula.  The regions used for the extraction of the X-ray spectra are shown.  The spectra from the northern region shows weaker iron L-shell emission near 1 keV, although both spectra match very well at very low and very high energies. Adapted from \citet{2007PASJ...59S.151H}.}
  \label{fig:neb}
\end{figure*}

\section{The Outer Debris}

Aside from reflected emission there's no evidence that the Homunculus itself is associated with any intrinsic emission produced by shock heating  \citep[although the ``constant X-ray emission component'' discovered by][ might show the effects of shock heating in the part of the Homunculus closest to \ec]{kenji}.  This is consistent with the Homunculus expanding into a nearly evacuated region.  However, substantial soft X-ray emission is associated with the so-called ``Outer Debris Field'', as shown in Figure \ref{fig:ecxo}.
The ``Outer Debris Field'' consists of nitrogen-rich material ejected in the giant eruptions of the 19$^{th}$ century, although some of the distant condensations may have originated  far earlier \citep{1978ApJ...219..498W} unless they were decelerated \citep{2005ASPC..332..271W}.

This outer X-ray emission is believed to be a ring rather than a limb-brightened shell \citep{2001ApJ...553..832S}, based on  imaging with the \chandra\ Observatory.  \citet{2004A&A...415..595W} compared \chandra\ X-ray images and spectra in the outer field with radial velocities from echelle spectra obtained at CTIO.  The X-ray emission in various places in the outer field had a temperature of $\sim 0.6$ keV, implying pre-shock gas velocities of $600-700$ km s$^{-1}$.  These velocities were in the general range of the optical velocities, although localized regions showed velocities of $1000-2000$ km s$^{-1}$.  Unlike the lobes of the Homunculus, the ``equatorial disk'' does seem to be a source of X-ray emission. \chandra\ images show a ``bridge'' of X-ray emission stretching across the outer field and apparently associated with the leading edge of the ``equatorial disk'' \citep[][ see figure \ref{fig:cartoon}]{2004A&A...415..595W}.  

\section{X-rays from the Carina Nebula}

The Carina Nebula itself is associated with soft, diffuse X-ray emission, as shown in Figure \ref{fig:neb}.  This emission has been known since the late 1970's \citep{1979ApJ...234L..55S}, though its origin is unclear.  The ``diffuse'' emission could be unresolved emission from a population of faint point sources, perhaps pre-main sequence stars.  It has often been attributed to the interaction of the combined winds from the massive star population with nebular material \citep{1979ApJ...234L..55S}, though the morphology of the emission does not follow closely the spatial distribution of the massive stars.  \citet{2007PASJ...59S.151H} obtained X-ray spectra of the emission with the X-ray Imaging Spectrometers (XIS) on the \suzaku\ X-ray observatory.  After correcting for most of the contamination due to strong point sources,  \citet{2007PASJ...59S.151H} showed that the iron L-shell emission was stronger in the region to the south of \ec\ than in the region to the north of \ec, and suggested that this difference represents a real variation in the iron abundance through the nebula.   This abundance variation is very large (a factor of 2-4), compared to the expected gradient over the Carina Nebula \citep[$<1\%$ according to the model Galactic iron abundance gradient of][]{2007A&A...462..943C} and thus unlikely to be primordial.  Such a large abundance gradient could have been produced by  a supernova, though  a supernova in the Carina Nebula would be rather astounding: it would have to have been produced either by a  star  more massive than the original mass of \ec\ (which is believed to have been $\sim100$ M$_{\odot}$ or so) or from an earlier generation of massive stars.

\section{Summary}

X-ray emission is a useful diagnostic of the stellar-nebular feedback loop in the Carina nebula on a range of scales. On the smallest scale ($\sim$ a few AU from \ec), X-rays produced by a wind-wind collision between \ec\ and a massive companion star serve as a probe of wind-driven mass loss and of the cool thick nebulosity silhouetted against the X-ray emitting region. On a slightly larger scale ($\sim$ 0.1 pc) the interaction of ejecta from giant eruptions (either the famous mid-19$^{th}$ century eruptions, or even earlier) with pre-existing nebula (or nebula shaped by the main-sequence wind from \ec\ and the companion) produce extended X-ray emission which can be used as a probe of the dynamics of the ejecta, and of abundance.  More extended ($\sim$ 10's of pc) diffuse X-ray emission traces abundance variations and either shows a large-scale variation in iron abundance across the nebula much larger than the Galactic gradient, possibly indicative of a supernova explosion.  


\end{document}